# Hong Kong Baptist University

Department of Geography

GEOG 4017 Geographical Information Systems

2022-2023 Semester 2

Final Course Report

Instructor: Dr. Song Jun

Topic:

Integrating GIS into Hong Kong Secondary School Geography Curriculum

Student Name: Lai Yin Ching

Student ID: 20227485

Word Count: 3398

# Integrating GIS into Hong Kong Secondary School Geography Curriculum

**Abstract**:

Hong Kong' senior geography curriculum has included GIS since the early 2000s. However, GIS in secondary schools does not play a significant role in Hong Kong secondary geography education. Analyzing GIS benefits by literature review, it is believed that GIS should be included in both the senior and junior geography curriculum. Moreover, the literature review indicates that without clear instruction from the Hong Kong Education Bureau (EDB), low preparedness of Hong Kong geography teachers, and unsupportive attitudes from academia and textbook publishers, GIS cannot be implemented in secondary schools of Hong Kong. Therefore, suggestions are made for the EDB, geography teachers, academia and textbook publishers to facilitate GIS involvement in senior and junior geography curriculums. The EDB can develop clear guidelines for teachers, academia and textbook publishers' references, and offer student-centered GIS educational courses for teachers. It is important for teachers to be prepared for advanced GIS technology and to even learn along with students. Academics and textbook publishers can provide free GIS maps targeted at Hong Kong' junior and senior geography curriculums. Although the report provides brief information towards the GIS implementation in Hong Kong geography education, it can inspire new ideas from other scholars to facilitate the usage of GIS in Hong Kong secondary school geography teaching.

## 1. Introduction

GIS (Geographic Information System), one of the academic fields in geospatial technology, is a computer system for inputting, collecting, storing, processing, displaying, presenting, and analyzing spatial data (Lam et al, 2009). The combination of geographical information and computer technology has made GIS widely used in a variety of industries, from scientific investigations to city planning to education, for any purpose. GIS demand increases significantly in the era of artificial intelligence and technology.

**2. Research Gaps**

While other countries and regions, for instance the United Kingdom, Turkish and Kazakhstan, have actively involved GIS in their high schools' geography curriculum, the Hong Kong Education Bureau (EDB) seems not to have an active role in promoting GIS in high schools and investigating the essential and valuable skill into its high school curriculum. Although the EDB highly emphasizes spatial analysis in both junior and senior geography curriculum, there are no guidelines or explanations on how to integrate GIS in geography teaching. GIS is even not mentioned in the junior geography curriculum. In addition, I discovered that only a few scholars have investigated integrating GIS into Hong Kong high school geography curriculum in depth. Most scholars investigate students' and teacher attitudes towards GIS and its advantages. Despite some scholar papers relating to the integration of GIS into the high school geography curriculum, those scholar papers were published in the early 2000s, which the secondary school geography curriculum has changed from the Hong Kong Certificate of Education Examination (HKCEE) and Hong Kong Advanced Level Examination (HKAL) to Hong Kong Diploma of Secondary Education Examination (HKDSE).

**3. Research Topic**

Based on the research gap, I would like to investigate GIS integration into the Hong Kong Secondary School Geography Curriculum from three perspectives, including the Education Bureau, teachers, and academia. Following the literature review, we will analyze three perspectives from the Education Bureau, teachers, and academia, as well as textbook publishers, to find out how GIS can be integrated into geography teaching.

## 4. Literature Review

### 4.1 Introductory of Literature Review

A modern world is characterized by technological advancements, changing world conditions, and people' interconnectedness. All of those are largely focused on spatial relationships. Thus, GIS benefits should be demonstrable, which benefits students' learning and career pathways. Moreover, the second part of the literature review discusses the role of GIS in educational affairs in Hong Kong and other regions and countries. As part of the analysis part, possible suggestions will be provided for integrating GIS into Hong Kong' high school curriculum.

### 4.2 Developing Geographic Thinking

GIS can develop students' geographical thinking, which further consolidates geography knowledge. GIS emphasizes spatial analysis, which uses maps to visualize region patterns and distributions. Through map reading, all the information can be clearly outlined. Different interactive layers and categories can also be added to web-based or software-based maps (Demeuov et al, 2021). Students can tick or untick the selected topics for comparison, classification, and analysis, which allows them to learn geography patterns and distributions in an interactive way. For example, when teaching about deforestation and human activities. GIS can show where deforestation occurs, how human activities keep expanding, and how serious is global deforestation? Aside from reading those maps before and after deforestation, students also need to analyze their responses by themselves, rather than simply memorizing knowledge from textbooks, which encourages them to examine maps and data in relation to cause and effect (Walshe, 2018). In addition, GIS enables students to use their creativity to solve problems, and it facilitates their understanding of geography as well (Industry Focus, 2015). GIS can be incorporated into lessons as homework and projects to require students to provide solutions to landslides using GIS. Students can think outside the box. Combining with the knowledge they learnt and urban problems in their living environment, students can use GIS to demonstrate possible solutions (Stevens, 2021). For example, contributing masonry retaining walls and installing soil nails to solve landslides. This can facilitate geography creative thinking skills and put geography knowledge into practice, which further consolidates geography knowledge. Therefore, GIS can facilitate students geographic thinking and reinforce their knowledge.

**4.3 Increasing Motivation in Geography Lessons**

GIS motivates students to engage in and participate in geography lessons. Arthvini (2010) and Stevens (2021) agreed that GIS has positive effects on students' motivation, attitude, and interpersonal skills. As GIS is considered high-level educational software in high schools, a majority of high school students have not tried GIS before. Using GIS is a new way to view the world from their own perspective. Additionally, GIS questions can relate to students' living environment, bringing knowledge to practice. This can motivate students to learn geography and explore and investigate the whole world. Walshe (2018) suggests local places around the student' residence and school can be implemented in junior school. For instance, when teaching urban planning in junior geography, students can use GIS for identifying different types of land use around their homes and schools. They may be interested in the pattern of city development. Within their living circle, which motivates them to learn more about Hong Kong' city pattern, and even study geography in high schools and universities (Chiu, 2017). By using GIS to engage students in lessons, teachers can activate the learning atmosphere in classrooms and ignite the passion and enthusiasm for geography (Kemp et al, 1992). In turn, integrating GIS into education industry increases students' motivation to study geography.

**4.4 GIS Involvements in other countries and regions education system**

GIS is highly integrated into other countries' and regions' geography curricula. A third of the geography curriculum of the General Certificate of Secondary Geography (GCSG) in the United Kingdom is GIS-based (AQA, 2016). Using GIS, satellite images, and different types of geographic representations, students are required to analyse and interpret qualitative and quantitative data from primary and secondary sources for their field work. Paper-and-pencil exams are also required for geographical applications in England, including the application of GIS in reality and other forms of geography representation, such as satellite images. Chinese and Taiwanese high schools already include GIS in their geography curriculums besides the United Kingdom. As stated by the Ministry of Education of the People' Republic of China, the high school geography curriculum includes basic GIS concepts and application. Students are engaged in geographical learning by applying a student-centered approach to teaching geography through GIS (Lam et al, 2009). In addition, Tai Wan' high school geography curriculum (2018) includes an elective topic related to spatial analysis, aiming to develop the importance of spatial information skills. The curriculum also incorporates problem-based investigation, for instance making theme maps to cover particular geographical issues, provoking discussion among students. Other countries and regions have extensively incorporated GIS into their geography curriculum.

**4.5 GIS Education in Hong Kong High School Geography**

GIS involvement in Hong Kong is not as extensive as other countries and regions (Tse, 2005). Although GIS has been introduced into Hong Kong's Senior Secondary Geography curriculum since 2009, the EDB in Hong Kong does not provide guidelines and instruction in integrating GIS into the junior or senior secondary geography curriculum. 96% of secondary teachers studied GIS in their undergraduate studies and attended GIS courses organized by the EDB or local universities. However, only 32% of teachers had used GIS before (Lam et al, 2009). 3 sectors, the EDB, teachers, and academia and textbook publishers pose barriers to GIS implementation in Hong Kong geography curriculum.

A challenge is implementing GIS into high school curriculum without clear instructions from the EDB. The EDB stated that junior students should be able to apply GIS to fieldwork and drawing basic maps (EDB, 2022). Senior students not only need to achieve those skills, but they should also be able to make judgments through comparing, analyzing, combining, and evaluating geography data with GIS. However, the EDB does not explicitly mention integrating GIS into classroom teaching (EDB, 2022). Secondary school teachers and textbook publishers do not know how to integrate GIS into lessons without the EDB' clear guidelines. There are confusions among teachers and textbook publishers about high school GIS level requirements and their relationship with geography topics. In addition, the GIS courses organized by the EDB emphasize a skill-based approach in which teachers are viewed as being developed from a top-down perspective. Practical knowledge of teachers is not taken into account. Thus, teachers not only acquire the skills of operating GIS but not involve GIS into practical teaching, which will only have a temporary effect.

There is a lack of self-preparedness among teachers in terms of taking on the responsibility of promoting GIS in geography curriculum. Nearly all teachers have taken GIS courses before. However, many teachers reported they had forgotten most of the GIS content on those courses because of the low frequency of GIS usage in those courses (Lam et al, 2009). A GIS training course organized by the EDB is mostly skill-based, which only teaches teachers about the basics of GIS usage (Chui, 2017). There is no linkage and relationship between the curriculum topics. In return, no teaching experience or tips are provided to teachers for reference. The use of GIS in teaching creates a knowledge gap among teachers. This leads teachers to push back GIS in classroom teaching. Similarly, teachers have to explore the GIS software by themselves due to the lack of useful information provided by the EDB (Tse, 2005).

Due to numerous administrative duties, inconvenient GIS applications, and teaching duties at schools, teachers often consider exploring GIS software that takes sustainability time and even pushes GIS back to high schools.

Academics and textbook publishers also pose burden in promoting GIS to high schools. Basic GIS software, digital data, and supporting educational software are needed for using GIS in teaching. Specifically, there is limited digital data suitable for the high school geography curriculum. Digital data, including digital maps, were expensive and unaffordable for some schools (Lam et al, 2009). The majority of teachers over 35 years old have forgotten all the GIS knowledge they learned in their undergraduate and teacher training courses, so choosing and processing suitable digital maps for their teaching materials is difficult and time-consuming (Lam et al, 2009). Although academic and textbook publishers strongly encourage teachers to include GIS in teaching, no GIS-supported digital data is provided for teachers to use.

**5. Joint-Cooperation in GIS implementation in High Schools**

To fully integrate GIS into the high school curriculum, the EDB, teachers, and academia and textbook publishers should work together to facilitate GIS implementation in secondary geography. The EDB should change the teaching method of a GIS training course from a skill-based approach to a student-centered approach and provide more digital maps for teachers. Teachers should promote learning in context and integrate GIS into their curriculum by applying a growth model that emphasizes learning in context. Academia and textbook publishers should raise general awareness of GIS by organizing GIS information days. They can facilitate GIS implementation in geographical teaching through designing suitable levels of digital maps based on geographical curriculum. The following paragraphs explain the above suggestions in detail.

## 5.1 EDB: Policy Formation and Educational Material Providers

Implementation of GIS in learning is hindered by the unclear role of GIS in the curriculum and the inappropriate teaching pedagogy of GIS training courses. Therefore, the EDB should develop policies for integrating GIS into high school geography classes and provide guidelines for teachers.

GIS, in the form of a session, can be incorporated into topics. Topic 2, 'Managing River and Coastal Environment: A Continuing Challenge' stated by the EDB as an example, an individual session can be included in the chapter for students' readings. In the GIS session, students have to analyze the landforms at the upper, middle, and lower courses of the river, and suggest examples of certain landforms around the world using GIS Story Maps by 'whys of where' (Kerski, 2015). For example, what types of terrain and features can be found on the upper course, middle course, and lower course of the river? How do they affect region relief? How do those landforms form? Why does flooding happen in those regions? All questions investigated in the GIS session should be related to the six themes and issues within the geographical curriculum (Figure 1). Steps for demonstrating the use of GIS in teaching geography should also be provided to teachers for reference and review. At the meantime, supported teaching materials should also be created by the EDB. With the leading role of the EDB and a clear standard set up by the EDB, teachers have a clear understanding of integrating GIS into teaching.

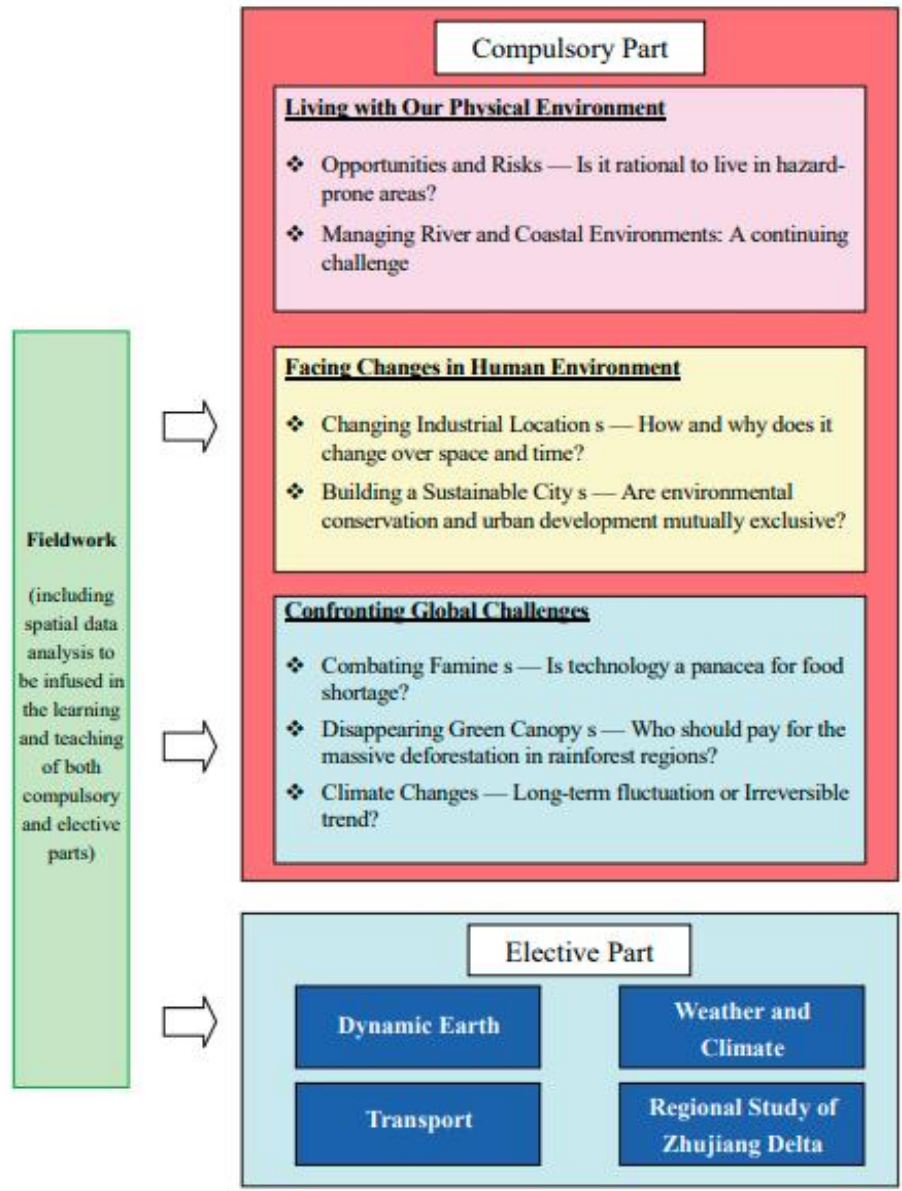

Figure 1: The Six Themes and Topics in the Hong Kong senior geography curriculum.

The GIS session should be related to the above 6 themes and topics in the curriculum.

Furthermore, the EDB might need to change its content from a skill-based approach to a student-centered approach. Instead of solely introducing spatial maps and educating teachers on GIS importance, teaching courses should provide demonstrations and examples of integrating GIS into geography teaching. For example, geography presentation and sharing can be done by Arc GIS Story Maps. This demonstration demonstrates understanding of the spatial distribution of industries around the world in theme 3, Changing Industrial Locations - How and Why Does It Change Over Time? A wide variety of GIS training courses should be designed for classroom design, and they emphasize the application of GIS to teaching geography in practice (Tse, 2005). The GIS knowledge and skills developed by teachers are connected in teaching. This, the role of GIS in the geography curriculum is clear, which facilitates the implementation of GIS into geographical curriculum.

**5.2 Teachers: Learning New Skills with Students**

Almost all geography teachers believe that familiarizing themselves with GIS software takes ample time, which makes it difficult for them to incorporate GIS into their teaching curriculum. However, geography teachers should try their efforts to acquire GIS knowledge and incorporate it into their classrooms. Teachers can access 15-minute *GeoInquiries* (Figure 2), which provide a wide range of human geography and physical geography, along with GIS digital maps (Tse, 2005). Moreover, there are many free GIS tutorial videos with step-by-step instructions on the Internet, namely *The GIS Hub* and *GIS & RS Solution* on YouTube (Figures 3&4). Following clear and direct instructions, teachers can acquire basic GIS skills within a couple of days and apply the newly learned skills to practice. Especially when teaching Regional Studies in the Zhujing Delta, an elective chapter in senior geography, teachers can incorporate GIS as educational activities. Most students find it hard to remember the Greater Bay Area cities. They are also confused about the economic roles of cities within the Greater Bay Area. The Arc GIS website I developed primarily introduces the economic activities development within the Greater Bay Area (Figure 5). By clicking into different cities in the digital maps, students can learn about the related knowledge, which contributes to the inclusion of GIS into classrooms, converting teacher-centered lessons into student-centered lessons, and increasing students' engagement. Furthermore, teachers can adopt a growth model of teaching, which emphasizes learning in context and acquiring new skills with students (Lam et al, 2009). After assigning GIS homework, teachers can do their own products for sharing and practicing their own GIS skills. Based on the above suggestions, teachers can become GIS promotors at schools.

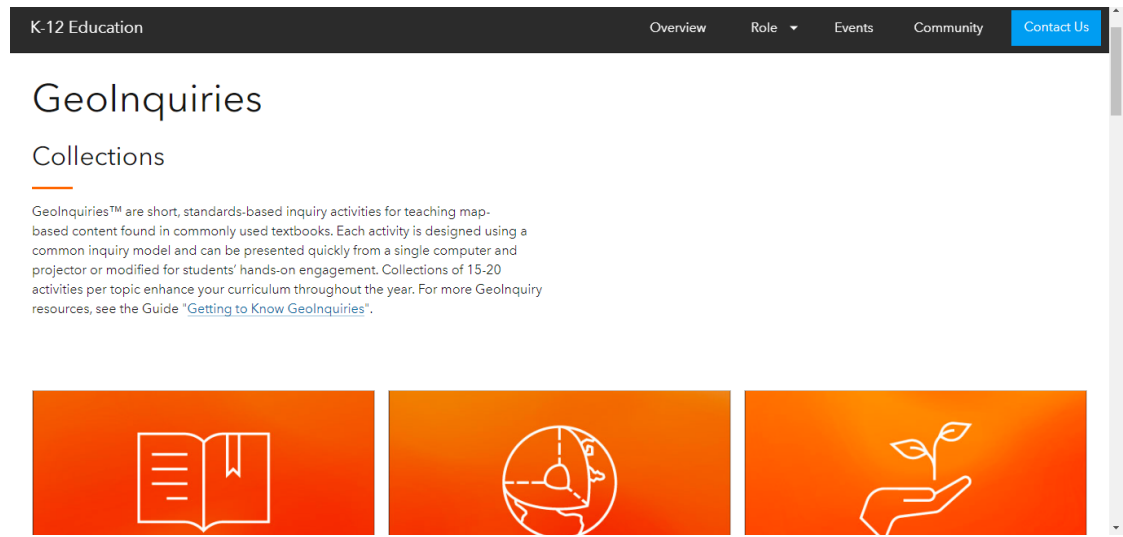
Figure 2: GeoInquires provides lots of GIS instructional activities and videos for teaching.

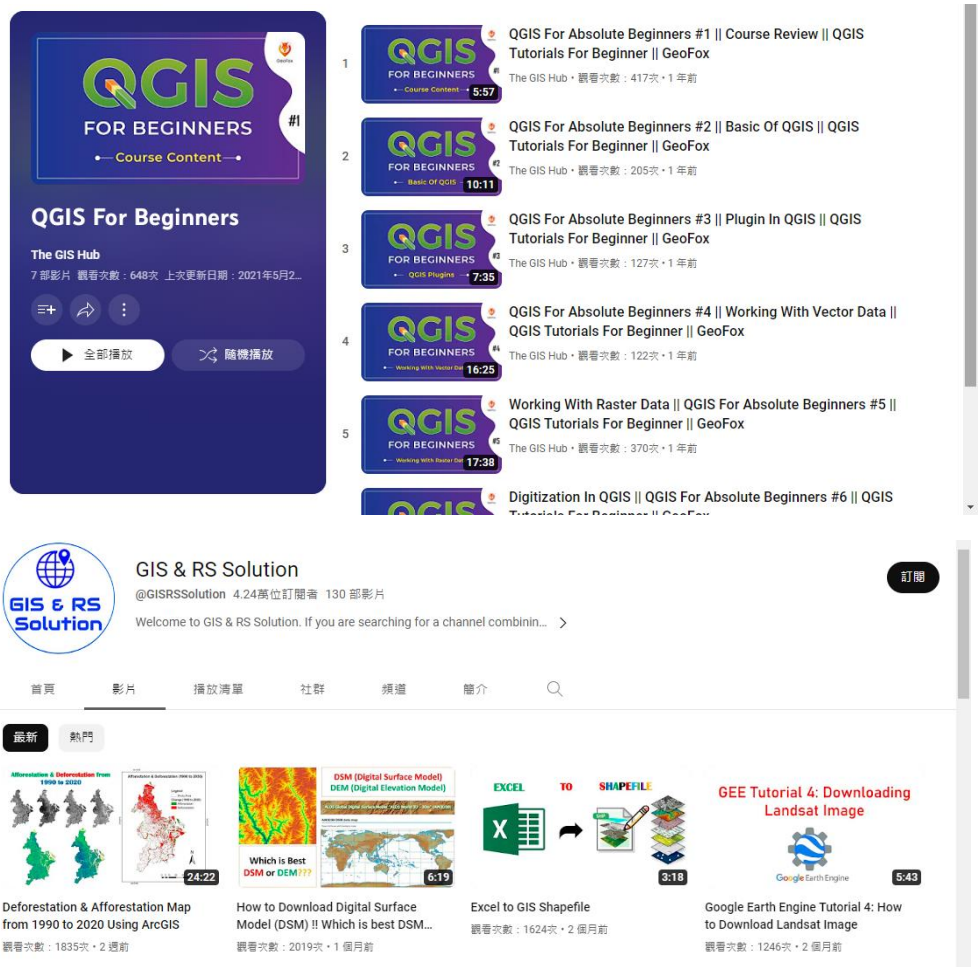
Figures 3&4: A list of free GIS tutorial videos are available in the Internet.

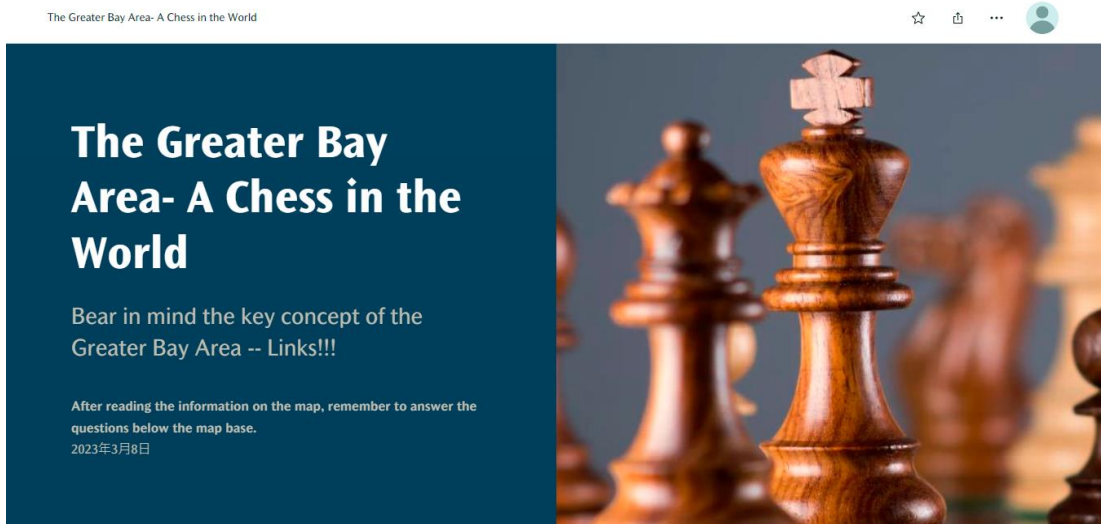

Figure 5: GIS can be used as educational activities in lessons.

Source from self created website, https://arcg.is/1GnX9j

## 5.3 Academia and Textbook Publishers: GIS Promoter and GIS Resource Provider

Referring to scholar papers, academia and textbook publishers do not actively involve GIS in the curriculum. None of GIS promotional activity targeting secondary schools and supporting educational digital data and maps has been organized and published respectively since GIS inclusion into Hong Kong geography curriculum in 2000. Hence, textbook publishers and academia should promote GIS in the geography curriculum by organizing GIS information days and providing teaching materials on GIS.

Organizing GIS information can increase students' and teachers' understanding of GIS. Referring to the past record, the first GIS day was held in 2002 by Esri China (Hong Kong) (Tse, 2005). The GIS day demonstrated practical application of GIS software and linkages with geography curriculum. However, the GIS day has shifted its focus to tertiary students since 2004. A similar situation can also be found in GIS competitions. Both events in GIS exclude secondary education. Esri China (Hong Kong) organizes the Esri Young Scholar Award every year for practicing GIS, but it is meant primarily for tertiary students. Therefore, GIS and textbook publishers can collaborate to organize GIS days and GIS competitions targeting secondary education. For instance, Intelligence CAD/CAM Technology Ltd (ICT), a Hong Kong company focusing on 3D technologies, organized 3D SOLIDWORKS Design Contest for secondary school students and tertiary school students to promote 3D printing technology in Hong Kong. Academia and textbook publishers might learn from ICT to organize events and competitions for secondary school students.

Likewise, textbook publishers and academia can develop GIS educational digital maps and platforms free for teachers and students. While the government' Geoinfro map, the Hong Kong Geodata Store provided by the Land Department, and Esri China (Hong Kong) Ltd' open Geospatial data are free, none of them are meant for secondary school teaching. For illustration, theme 1 in the curriculum, 'Opportunities and Risks – is it rational to live in hazard prone areas?', academia and textbook publishers can create map packages relating to the spatial relationships of mountains, island arcs, ocean trenches, volcanic activities and earthquakes. Supplementary worksheets can also be designed to consolidate students' knowledge. With sufficient teaching materials, teachers and students will be able to integrate GIS seamlessly into normal teaching, which will not pose burdens in the implementation of GIS.

**6. Conclusion**

We live in a constantly changing world. A variety of physical changes and human activities change our daily lives, including volcanic eruptions, shifting plates, and urbanization. GIS not only beneficial to students learning geography, including fostering students' geographical thinking and increasing students' learning motivation in geography, but it is also the core skills of studying geography. Geography students should be able to apply GIS appliances to analyze spatial changes, patterns, and relationships geographical data. The EDB, teachers, academia, and textbook publishers should work hand-in-hand to fully integrate GIS into the current geography curriculum. EDB should begin by providing clear instructions on including GIS into teaching, then update its teaching method in GIS training courses to a student-centered approach and provide more teaching examples. Teachers are also suggested to explore GIS software through online videos and use the growth model in their lessons. In the same way, academics and textbook publishers should provide GIS education platforms for secondary geography students and organize events for secondary students. The paper is intended to serve as a catalyst for GIS in Hong Kong secondary geography teaching and to stimulate new ideas. It is hoped that GIS will be recognized in Hong Kong education to empower students to change the world positively.

Word Counts: 3398

*國民中小學暨普通型高級中等學校*. https://cutt.ly/z6Q8How